\documentstyle[preprint,aps,eqsecnum]{revtex} 
\tighten
\begin{document}
\draft
\title{A New Theory of Stochastic Inflation}
\author{ Andrew Matacz
\thanks{e-mail address: andrewm@maths.su.oz.au}}
\address{School of Mathematics and Statistics \\
University of Sydney\\ NSW, 2006, Australia}
\maketitle
\begin{abstract}
The stochastic inflation program is a framework 
for understanding the dynamics of a 
quantum scalar field driving an inflationary phase. 
Though widely used and accepted, there have over recent years 
been serious criticisms of this theory. 
In this paper I will present a new theory of stochastic 
inflation which avoids the problems of the conventional 
approach. Specifically, the theory can address the 
quantum-to-classical transition problem, and it will be shown 
to lead to a dramatic easing of the fine tuning 
constraints that have plagued inflation theories.
\end{abstract}
\pacs{98.80.Cq, 05.40.+j, 03.70.+k}
\newpage
\section{Introduction}

The inflationary universe scenario asserts that, at some very early 
time, 
the universe went through a de Sitter phase expansion with scale 
factor $a(t)$ growing as $e^{Ht}$. Inflation is 
needed because it solves the horizon, flatness and monopole problems 
of the very early universe
and also provides a mechanism for the creation of 
primordial density fluctuations. For these reasons it is an integral 
part of the standard cosmological model 
\cite{inflation}. 

The inflationary phase is driven by a quantum 
scalar field with a potential $V(\Phi)$, that can take on many 
different 
forms that satisfy the `slow roll' conditions. Inflationary 
scenarios fall into two  
types. In the first, which includes the so-called
old and new inflation, the scalar field driving inflation 
(the inflaton) is assumed to be in thermal equilibrium 
with the rest of the universe. The universe is assumed to obey the 
standard hot big bang cosmology in the 
period preceding and after inflation. In the chaotic inflation 
scenario these 
assumptions are relaxed. 
Instead the inflaton is assumed to be only very weakly coupled to 
other fields. 
This makes it possible to choose initial states for the inflaton 
which are far from equilibrium. 
The standard big bang cosmology now only applies after the reheating 
stage of inflation.

In the conventional approach to inflaton dynamics \cite{inflation}, 
the inflaton field $\Phi$ is first 
split into a homogeneous piece and an inhomogeneous piece
\begin{equation}
\Phi({\bf x},s)=\phi(s)+\psi({\bf x},s),
\end{equation}
where $\phi$ is interpreted as the field $\Phi$ coarse-grained over 
a volume 
$\Omega$:
\begin{equation}
\phi(s)=\frac{1}{\Omega}\int_{\Omega}\Phi({\bf x},s)d^3{\bf x}.
\end{equation} 
In this description all information on scales 
smaller than the coarse-graining volume, such as density 
fluctuations, will be contained in $\psi$.
This means the field $\psi$ has a low frequency cutoff in its 
spectrum. 
The coarse-graining volume is set by the 
causal horizon which leads to $\Omega\sim H^{-3}$. We will refer to 
this $\phi$ as 
the global order parameter since it is constructed from a 
coarse-graining volume that is always 
larger than the observable universe. 
The dynamics of the global order parameter is then postulated to 
obey 
the classical `slow roll' equation of motion 
\begin{equation}
\dot{\phi}+\frac{V'(\phi)}{3H}=0.
\end{equation} 
This equation governs the dynamics of $\phi$ which drives the 
inflationary phase.
It is also possible to discuss the generation of primordial density 
fluctuations using $\psi$.
Assuming that $\phi\gg \psi$, 
it can be shown that $\psi$ is described by a free massless 
minimally 
coupled quantum scalar field. During exponential inflation the 
quantum fluctuations 
of $\psi$ grow as $\langle\psi^2\rangle\sim H^3t$. These quantum 
fluctuations are then 
identified with the classical 
fluctuations which generate primordial density fluctuations 
\cite{inflation,fluct}.

Although the conventional theory sketched above is 
widely accepted, it must be considered problematic for three 
important  
reasons:
\begin{itemize}
\item it is assumed, without justification, that the 
global order parameter $\phi$ can be treated as a classical 
order parameter, and that the quantum fluctuations of 
$\psi$ are equivalent to classical fluctuations.

\item the fluctuations of $\psi$ lead to an overproduction of 
primordial density fluctuations. This can 
only be avoided by unnaturally fine tuning the coupling constants in 
the inflaton potential. 

\item the backreaction of $\psi$ on the dynamics of $\phi$ is 
ignored.
\end{itemize}
Based on the conventional theory there should be no reason to expect 
$\phi$ and $\psi$ 
to behave classically. 
The action for $\phi$ is non-linear and strongly time dependent. 
Both of these properties will always generate 
quantum mechanical coherence over time.  
Similarly, because $\psi$ is treated as a free quantum field 
its quantum vacuum state will always remain spatially homogeneous. 
It is only when these fluctuations become classical that spatial 
inhomogeneity 
is generated. 
Since the conventional theory treats
$\phi$ and $\psi$ as independent closed systems 
it is impossible in principle for it to explain 
the 
quantum-to-classical transition of $\phi$ and $\psi$. The purpose of 
this paper 
is to develop a more considered approach to inflaton dynamics which 
can 
address these three critical issues.

The problems outlined above can all be addressed by a 
careful application of the principles and techniques of 
non-equilibrium 
quantum statistical physics. 
When we coarse-grain the scalar field as in equation (1.2), what we 
are effectively doing is splitting our closed 
quantum system $\Phi$ into a system sector $\phi$ and an environment 
sector $\psi$. 
The system sector includes only wavelengths long 
compared to the coarse-graining scale while the environment 
sector comprises wavelengths short compared to the coarse-graining 
scale. 
By discarding information about the environment sector $\psi$, 
and assuming that the system and environment interact,
we turn the unitary, reversible dynamics of the closed system 
$\Phi$, into the nonunitary and irreversible 
dynamics of an open system $\phi$. Thus we have 
gone from a fundamental theory to an effective theory. 
The backreaction of any
environment sector on a system sector will generally manifest itself 
by introducing noise, 
dissipation and renormalization effects into the dynamics of the 
system \cite{qos}.
The noise and dissipation will be related by some 
fluctuation-dissipation relation. 
Noise will induce diffusion and decoherence (the loss of 
quantum mechanical coherence). Decoherence is the 
critical ingredient if we are to dynamically 
demonstrate the quantum-to-classical transition 
of an open system. 

In this approach our only interest in the environment $\psi$  
is its backreaction on the system dynamics. In this case the 
global order parameter is no longer a suitable system variable
because it contains no information about spatial structure within 
our universe.
A system variable that does contain this information can be 
obtained simply by choosing the apparent rather 
than the causal horizon as the coarse-graining scale. This 
corresponds to a coarse-graining 
volume of $\Omega\sim (He^{Ht})^{-3}$. 
With this averaging volume we  will refer to $\phi$ 
as the local order parameter. 
The use of a local rather than global order parameter for inflaton 
dynamics has been strongly advocated by Morikawa \cite{mm}.
In this scheme it is the classical fluctuations $\delta\phi$ of the 
local order parameter which lead to 
density fluctuations rather than the quantum fluctuations derived 
from $\langle\psi^2\rangle$. The new role of the field $\psi$ is to 
provide a noise source (via backreaction) in the quantum dynamics 
of the local order parameter. This noise will generate quantum 
decoherence which is the process that creates the classical 
fluctuations $\delta\phi$ of the local order parameter. 
This fundamental conceptual shift regarding the role of 
$\psi$ is the key to developing a theory that can address the 
quantum-to-classical transition problem. As a bonus it will be 
shown that the new theory leads to a dramatic easing of the 
fine tuning constraints that plague the conventional approach 
to inflaton dynamics.

A first attempt to study the backreaction of $\psi$ on the local 
order parameter $\phi$, can be found in the 
`stochastic inflation' program initiated by Starobinsky \cite{staro} 
and further developed by others \cite{stoinf}.
This program claims that the semiclassical equation of motion 
for the local order parameter is given by the Starobinsky equation
\begin{equation}
\dot \phi + \frac{V'(\phi)}{3H} = \frac{H^{3/2}}{2\pi}F_w (t),  
\end{equation}
where $F_w(t)$ is a zero mean gaussian white noise source of unit 
amplitude.
The system field equation is thus transformed into a classical 
Langevin equation with a white noise source.
A Fokker-Planck equation can also be derived which depicts
the evolution of the probability distribution of the scalar field. 
Since it was first derived the Starobinsky equation 
has stimulated numerous studies.
Much effort has been devoted to the solution of this stochastic 
equation for a description of the 
inflationary transition and the generation of primordial density 
fluctuations \cite{sifluct}. The same equation forms the 
theoretical foundation for studies on the very large scale 
structure of the 
universe \cite{vlss}. It has been  
claimed that the stochastic inflation program can explain the 
quantum-to-classical transition of the coarse-grained field 
\cite{qct}. 
A quantum mechanical 
description, for which the Starobinsky equation is the semiclassical 
limit, has been developed using methods to quantize 
classically dissipative systems 
\cite{graz} (this list of references is by no means complete).

In the derivation of the Starobinsky equation, interactions 
between the local order parameter $\phi$ and its short wavelength 
environment $\psi$ are derived from the quadratic terms in 
the scalar field action. Interactions that derive from 
a self-interacting potential are assumed to be small and are 
neglected. 
This procedure cannot be justified because, for free fields,  
there is no mode-mode coupling and therefore no way for 
short wavelengths to backreact on long wavelengths. 
Mode-mode coupling can only be generated by a self-interacting 
potential.
Advocates of stochastic inflation would say that the coupling is 
generated by the time dependent nature of the 
system environment split. 
However it is simply impossible for a physical coupling to 
be generated by a time dependent 
choice of what constitutes the system and environment 
\cite{cgea,HuBelgium,CH1}. We would 
therefore expect there to be, as long 
as self-interaction is ignored, no coupling and hence no 
backreaction of the environment $\psi$ on the local order 
parameter $\phi$. This implies that classical fluctuations 
in the local order parameter $\phi$ must vanish for a free 
theory. We therefore conclude that the Starobinsky equation 
fails to describe the backreaction of $\psi$ on the 
local order parameter $\phi$.

A better interpretation for the Starobinsky equation is that it 
models the conventional 
approach to inflaton dynamics with a classical stochastic 
dynamical system. The left hand side is the usual postulated 
slow roll equation of motion for the global order parameter. 
The stochastic term is meant to {\it classically} model the quantum 
fluctuations of $\psi$.
This interpretation is plausible because solving the free 
Starobinsky 
equation gives $\langle\phi^2\rangle\sim H^3t$. 
This is the same as the 
the growth of quantum fluctuations in $\psi$ in the conventional 
approach.
However, since it is essentially equivalent to the conventional 
approach, Starobinsky's stochastic inflation program  
suffers from the same serious problems that 
were previously outlined. 

Several authors 
have previously made similar criticisms of Starobinsky's stochastic 
inflation and the conventional inflaton dynamics.
Hu and Zhang \cite{cgea} first questioned the ability of a 
time-dependent split in generating noise for a free field, and 
introduced the coarse-grained effective action for a proper 
treatment of backreaction. 
Habib \cite{Hab} demonstrated that the Starobinsky theory of 
stochastic inflation does not 
address the quantum-to-classical transition despite claims that it 
does \cite{qct}. Hu, Paz and Zhang 
\cite{HuBelgium} (see also \cite{HuWaseda,HuTsukuba}) provided an 
extensive critique of the Starobinsky 
theory and developed their own theory of stochastic inflation for 
the case of a quartic potential. Morikawa \cite{morwas} 
pointed out that self-interaction is necessary for the 
generation of fluctuations, and that a lack of 
understanding of this point is the cause of the fine tuning 
problem of inflation. This point has 
also been emphasized by Calzetta and Hu \cite{CH1} in a 
recent discussion on the problems with conventional inflaton 
dynamics. See also Morikawa \cite{mm} for a general discussion 
on inflaton dynamics.

Clearly a new theory of stochastic inflation is required which can 
address the 3 problems of the conventional 
approach to inflaton dynamics. This new theory will require 
a fully nonequilibrium quantum mechanical formalism which deals 
with the statistical nature of mixed states and the dynamics of 
reduced
density matrices. An ideal formalism for this is the 
Feynman Vernon influence functional \cite{FeyVer}. The 
influence functional describes the averaged effect of the 
environmental degrees of freedom on the 
system degrees of freedom to which they are coupled. With the 
influence functional we can unambiguously identify a noise and 
dissipation 
kernel related by some generalized fluctuation-dissipation 
relation. 
The formalism leads directly to a propagator for the reduced 
density matrix which can then be used 
to derive a master equation. The propagator can be used to study 
the decoherence process which is the critical part of the 
quantum-to-classical transition. The 
master equation can be transformed into a Fokker-Planck type 
equation for 
the Wigner function. In the semiclassical limit we have a 
generalized Langevin 
equation.

The stochastic inflation program, and indeed many other problems 
in the early 
universe \cite{HuWaseda}, fall under the paradigm of 
quantum Brownian motion (QBM). QBM is 
one of the two major paradigms of 
non-equilibrium quantum statistical mechanics which is 
amenable to detailed analysis (the other being Boltzmann's 
kinetic theory).
The complexity of the problems in cosmology led Hu, Paz and Zhang 
\cite{HPZ} to first consider a systematic study of QBM in a 
general environment using
the influence functional formalism. The special features
associated with a nonohmic bath, or ohmic bath at low 
temperatures are the appearance of colored noise and nonlocal 
dissipation. The methodology and viewpoint of QBM have been
applied to the analysis of some basic issues in quantum cosmology
\cite{HuErice,HuTsukuba,Sinha,PazSin,HPS,decQC},
effective field theory \cite{HuPhysica,HuBanff},
and the foundation of quantum mechanics, such as the uncertainty 
principle
\cite{HuZhaUncer,AndHal} and, most significantly, decoherence
\cite{envdec,conhis,GelHar1,GelHar2,CalHuDCH}
in the quantum-to-classical transition problem. (See the  reviews of
\cite{ZurekPTP,HarLH,Omnes} and references therein for the standard 
literature on this topic). The 
QBM models in \cite{HPZ} were further generalized by Hu and Matacz 
\cite{HuMat} by making the system and bath oscillators the most 
general time 
dependent quadratic oscillators, and by considering more general 
system-bath couplings. 
The time dependence generates parametric amplification (squeezing) 
in the oscillators which is precisely how 
quantum fluctuations of a free field are amplified in the 
inflationary universe \cite{Mat}.
This formalism allows for a general study of nonequilibrium quantum 
statistical processes in time dependent backgrounds, as is the case 
in the early universe. 

Using the same methodology as in the previous QBM studies, Hu, Paz 
and Zhang 
\cite{HuBelgium} were the first, based on the system environment 
split orginally proposed by Starobinsky, to correctly formulate 
the problem of the backreaction of short wavelength quantum 
fluctuations $\psi$, on the local order parameter $\phi$. For more 
recent work of a similar nature see Lombardo and Mazzitelli 
\cite{sim}. These works are based on a rigorous quantum 
field theory of open systems, however  they do have some practical 
shortcomings. The dynamical systems obtained are of 
a functional nature further complicated by non-local 
dissipation and colored noise. It is very difficult to 
analyse the quantum and semiclassical dynamics of these systems. 
The models are restricted to flat space or conformally 
coupled fields in a de Sitter phase. For applications to inflation 
we are mainly interested in minimally coupled scalar fields 
in a de Sitter phase.
The models are very specific, involving a $\lambda\Phi^4$ 
self-interaction. Deriving the renormalized effective action for 
the long wavelength sector is a complex calculation that can only 
be done in the context of specific potentials. In 
inflation there is great interest in a wide variety of 
potentials.
In this case the quantum field theory calculations may be much more 
difficult to implement than for the $\lambda\Phi^4$ case. 
Because of these problems there has not yet been any detailed 
investigation of inflaton dynamics based on these quantum field 
theoretic open systems.   

In order to proceed further we need to develop a new simplified 
theory that overcomes these practical problems.          
This must be done without compromising a rigorous treatment based 
firmly on the principles of 
non-equilibrium quantum statistical physics. Only a theory of this 
nature can address the failures of conventional inflaton 
dynamics. In this paper such a theory is developed. 
The main result is that, for a minimally coupled scalar field 
in a de Sitter phase, the quantum dynamics of the local 
order parameter $\phi$ can be described by the relatively 
simple stochastic quantum mechanical Hamiltonian 
\begin{equation}
H(t)=\frac{p^2}{2e^{3Ht}}+e^{3Ht}V(\phi)-\frac{H^2}
{8\pi^3}e^{3Ht}~V''(\phi)F_c(t),
\end{equation}
where $p=e^{3Ht}\dot{\phi}$ is the canonical momentum and $p$ and 
$\phi$ obey the usual quantum mechanical commutator. 
$F_c(t)$ is a zero mean gaussian colored noise of unit amplitude
with a correlation time of the order $H^{-1}$. 
This result is valid for a general inflaton potential.
The origin 
of the noise is the backreaction of 
quantum fluctuations with wavelengths shorter than the 
coarse-graining scale. The noise is of a multiplicative nature 
because its origin is the mode-mode coupling 
induced by the self-interaction of the inflaton. For a free 
field this theory predicts no noise term. This should be expected
from arguments made here and previously 
\cite{cgea,HuBelgium,morwas,CH1}. 
The major simplification is made by ignoring information about 
spatial correlations between the order parameters of different 
regions. This allows a description based on a single degree of 
freedom. Further significant simplification is obtained by 
invoking the standard slow roll assumptions. This makes it 
possible to show that the potential 
renormalization and non-local dissipation terms are negligible.

The approximate semiclassical limit of the 
quantum open system (1.5) is 
\begin{equation}
\dot{\phi}+\frac{V'(\phi)}{3H}
=\frac{H^{1/2}}{\sqrt{864}\pi^3}V'''(\phi)F_w(t).
\end{equation}
In deriving this equation the slow roll assumptions we used to 
neglect the inertial term and approximate the colored noise 
by a white noise $F_w$.
This equation is no more complicated than 
the widely used Starobinsky equation (1.4). 
However it fundamentally 
differs from it because the noise term vanishes for a free theory.
Equation (1.6) is a superior alternative to the 
Starobinsky equation because
it is the semiclassical limit of the quantum open system (1.5). 
This allows the validity of the semiclassical limit to be 
dynamically derived by investigating the quantum decoherence 
generated by the noise in (1.5) 
(On the other hand, the `classicality' of the Starobinsky equation 
is simply postulated and is not a natural outcome of any 
quantum open system). 
The other crucial advantage is that, as we will see, 
equation (1.6) leads to a dramatic easing of the fine 
tuning constraints.

\section{Stochastic Inflation and Quantum Brownian 
Motion}
In this section we will see how the dynamics of a 
coarse-grained scalar field in an expanding universe, can be 
described in terms 
of time dependent quantum Brownian motion with non-linear
system-bath coupling. 
We consider a minimally coupled scalar field 
evolving in a spatially flat background with the metric
\begin{equation}
ds^2=dt^2-a^2(t)d^2{\bf x}.
\end{equation}
The action has the form
\begin{equation}
S=\int_{t_i}^{t}ds\int_{\Omega (s)}d^3{\bf x}~{\cal L}({\bf x},s)
\end{equation}
where
\begin{equation}
{\cal L}({\bf x},s)=a^3(s)\left[\frac{1}{2}\dot{\Phi}^2({\bf x},s)-
\frac{\Bigl(\bigtriangledown\Phi({\bf x},s)\Bigr)^2}{2a^2(s)}
-V\Bigl(\Phi({\bf x},s)\Bigr)\right].
\end{equation}
We assume $t_i$ is our initial time and $a(t_i)=1$. 

As is usual in inflation, we split $\Phi$ into a homogeneous piece 
and an
inhomogeneous piece
\begin{equation}
\Phi({\bf x},s)=\phi(s)+\psi({\bf x},s),
\end{equation}
and interpret $\phi$ as the field $\Phi$ coarse-grained over a 
volume $\Omega(s)$. We therefore write
\begin{equation}
\phi(s)=\frac{1}{\Omega(s)}\int_{\Omega(s)}\Phi({\bf x},s)d^3{\bf x}
\end{equation} 
which will be true as long as
\begin{equation}
\int_{\Omega(s)}d^3{\bf x}~\psi({\bf x},s)=
\int_{\Omega(s)}d^3{\bf x}~\dot{\psi}({\bf x},s)=0.
\end{equation}
Assuming $\phi\gg\psi$, the potential term in (2.3) becomes
\begin{equation}
V(\Phi)\simeq V(\phi)+\psi V'(\phi)+\frac{\psi^2}{2}V''(\phi).
\end{equation} 
Substituting (2.7) into (2.3) and using (2.6) we find
\begin{eqnarray}
S&=&\int^t_{t_i} ds~ \Omega (s)a^3(s)\left(\frac{1}{2}\dot{\phi}^2(s)
-V(\phi)\right)
-\frac{1}{2}\int^t_{t_i} ds~ a^3(s) V''(\phi)\left(\int_{\Omega (s)}
d^3{\bf x}
~\psi^2({\bf x},s)\right) \nonumber \\
&+&\int^t_{t_i}ds\int_{\Omega (s)}d^3{\bf x}~a^3(s)
\left[\frac{1}{2}\dot{\psi}^2({\bf x},s)-
\frac{\Bigl(\bigtriangledown\psi({\bf x},s)\Bigr)^2}{2a^2(s)}\right].
\end{eqnarray}
The coarse-graining breaks the scalar field up into cells of volume 
$\Omega$, each 
labeled by a coarse-grained position variable. We have dropped all 
reference to 
the position label on $\phi$ since the problem is spatially 
invariant. By 
dealing with only one degree of freedom we are ignoring information 
about 
spatial correlations between different cells. This is the reason why 
the 
middle term in the expansion (2.7) plays no role. Later in this 
section we justify 
why we can drop the spatial gradient term of the system sector.

The environment field $\psi$ can be written as
\begin{equation}
\psi({\bf x})=\sqrt{\frac{2}{\Omega(s)}}\sum_{{\bf k}}[q_{{\bf k}}^+ 
\cos {\bf k}
\cdot{\bf x}+q_{{\bf k}}^-\sin {\bf k}
\cdot{\bf x}]
\end{equation}
in which case the action (2.8) becomes
\begin{eqnarray}
S&=&\int^t_{t_i} ds~ \Omega (s)a^3(s)\left[\frac{1}{2}\dot{\phi}^2(s)
-V(\phi)\right]
-\frac{1}{2}\sum_{\sigma}^{+-}\sum_{{\bf k}}
\int^t_{t_i} ds~ a^3(s) V''(\phi(s))q_{{\bf k}}^{\sigma 2} \nonumber 
\\
&+&\frac{1}{2}\sum_{\sigma}^{+-}\sum_{{\bf k}}\int^t_{t_i}ds~a^3(s)
\left[(\dot{q}_{{\bf k}}^{\sigma})^2-
\frac{k^2}{a^2(s)}q_{{\bf k}}^{\sigma2}\right]
\end{eqnarray}
where $k=|{\bf k}|$. The action (2.10) is now in  
the form of a time-dependent one dimensional 
system, non-linearly coupled to an environment of time dependent 
harmonic oscillators. This action is a generalization 
of the well known quantum Brownian motion problem 
\cite{FeyVer,HPZ,HuMat,HuMat1}. 
In the next section we will show how the influence functional 
method can be used to describe the averaged effect of the 
environment on the quantum dynamics of the system.

\subsection{Low Frequency Cutoff}
We will impose a long wavelength cutoff in the field 
$\psi$, such that the longest wavelength equals the diameter of the 
coarse-graining volume. This is 
necessary since we are interpreting $\phi$ as the 
field $\Phi$ coarse-grained over the volume $\Omega(s)$.
This means that all frequencies less than the cutoff 
are considered part of the system sector. 
Since there is no mode-mode coupling for free fields, 
we cannot expect $\psi$ to couple to $\phi$ in this case. This is 
clearly true for the action (2.10). 
Writing the volume element as
\begin{equation}
\Omega(s)=\frac{4\pi^4}{3\epsilon^3\Lambda^3(s)}
\end{equation}
we can express the low frequency cutoff as 
\begin{equation}
k_{min}=\epsilon\Lambda(t)
\end{equation}
where $\epsilon$ is an arbitrary  dimensionless parameter that 
scales the coarse-graining volume.
An obvious question is how do we choose $k_{min}$? 

At a purely technical level we want 
to choose $\Omega$ large enough so that the spatial gradient 
term in (2.3) can be neglected for the system sector. 
The appropriate volume can be deduced from
the classical equation 
of motion for the environment modes.
Writing $Q_k=q_k/a(s)$ we find this equation is 
\begin{equation}
\partial_{\eta}^2Q_k + Q_k\Bigl(k^2-(\partial_{\eta}^2a)/a\Bigr)=0
\end{equation}
where $\eta=\int dt/a(t)$ is conformal time. We can see that the 
environment spectrum is split, 
with an unstable low frequency sector and a stable effectively flat 
space sector.
Choosing 
\begin{equation}
\Lambda(t)=\sqrt{(\partial_{\eta}^2a)/a},\;\;\;\epsilon\ll 1
\end{equation}
amounts to including all those modes where the spatial gradient is 
significant in the 
environment sector.

Detailed results in this paper will be limited to de-Sitter 
inflation where \newline 
$a(s)=e^{Hs}$, $\eta=-e^{-Hs}/H$ and $t_i=0$. In this case (2.14) 
and (2.12) lead to
\begin{equation}
k_{min}=\epsilon Ha(t),\;\;\;\epsilon\ll 1.
\end{equation}
This cutoff satisfies our technical requirement. We also have the 
physical requirement that information 
about large scale structure is contained in the system sector. 
This requires $\epsilon$ to be small but {\it finite}.
Taking $\epsilon\rightarrow 0$ would correspond to an over coarse-grained 
system that would contain no information of observable scales. 
The scale of the observable universe today 
is expected to cross the Hubble radius at least 7 e-foldings after 
inflation begins \cite{inflation}. This means 
that todays horizon scale maps to $k\sim 10^3 H$. We can see from 
(2.15) that after a few e-foldings, 
scales corresponding to our observable universe are shifted into 
the system sector. 
Of course, (2.15) is the same as Starobinsky's choice 
\cite{staro,stoinf} for the low frequency cutoff.

\section{Influence Functional}
Consider the quantum system described by the action
\begin{equation}
S[\phi,{\bf q}]=S[\phi]+S_e[{\bf q}]+S_{int}[\phi,{\bf q}].
\end{equation}
We will take $\phi$ as our system variable and ${\bf q}$ to be our
environmental variables. Typically the environment has infinite
degrees of freedom, denoted here by a bold type.
We will briefly review here the Feynman-Vernon
influence functional method for deriving the evolution operator 
\cite{FeyVer}.
The method provides an easy way to obtain a functional
representation for the evolution operator ${\cal J}_r$
of the reduced density matrix $ \hat\rho_r $.

We are interested in the reduced density matrix defined as
\begin{equation}
\rho_r(\phi,\phi')
=\int\limits_{-\infty}^{+\infty}d{\bf q}\int\limits_{-\infty}^
{+\infty}d{\bf q'}\rho(\phi,{\bf q};\phi',{\bf q}')
\delta({\bf q}-{\bf q}'),
\end{equation}
\noindent which is propagated in time by the evolution operator 
${\cal J}_r$
\begin{equation}
\rho_r(\phi,\phi',t)
=\int\limits_{-\infty}^{+\infty}d\phi_i\int\limits_{-\infty}^
{+\infty}d\phi'_i~
 {\cal J}_r(\phi,\phi',t~|~\phi_i,\phi'_i,t_i)~
\rho_r(\phi_i,\phi'_i,t_i~).
\end{equation}
By using the functional representation of the full density matrix 
evolution
operator given in (3.2), we can also represent ${\cal J}_r$ in path 
integral 
form.
In general, the expression is very complicated since the evolution 
operator
${\cal J}_r$ depends on the initial state. If we assume that at a 
given initial time $t_i$, the system and the environment 
are uncorrelated
\begin{equation}
\hat\rho(t_i)=\hat\rho_s(t_i)\times\hat\rho_e(t_i),
\end{equation}
then the evolution operator for the reduced density matrix does not 
depend
on the initial state of the system and can be written as
\begin{equation}
{\cal J}_r(\phi_f,\phi'_f,t~|~\phi_i,\phi'_i,t_i)
 =\int\limits_{\phi_i}^{\phi_f}D\phi
   \int\limits_{\phi'_i}^{\phi'_f}D\phi'~
   \exp\left\{i\Bigl\{S[\phi]-S[\phi']\Bigr\}\right\}~
{\cal F}[\phi,\phi']. 
\end{equation}
The factor ${\cal F}[\phi,\phi']$,  called the 
`influence functional', is defined as
\begin{eqnarray}
{\cal F}[\phi,\phi']&=&\exp\left\{i S_{IF}[\phi,\phi']\right\} 
\nonumber \\ 
&=& Tr\Bigl(\hat{U}[\phi_{t,t_i}]\hat{\rho}_e (t_i)
\hat{U}^{\dag}[\phi'_{t,t_i}]\Bigr),
\end{eqnarray}
where $\hat{U}$ is the quantum propagator for the action
$S_e[{\bf q}]+S_{int}[\phi(s),{\bf q}]$ with
$\phi(s)$ treated as a time dependent classical forcing term.
We have found this form to be very convenient for deriving the 
influence
functional \cite{HuMat,HuMat1}. $S_{IF}[\phi,\phi']$ is the 
influence action and the effective action  for the open quantum 
system is defined as
$S_{eff}[\phi,\phi'] = S[\phi]-S[\phi'] + S_{IF}[\phi,\phi']$.

If the interaction term is 
zero, then it is obvious from its definition, that the
influence functional is equal to unity and the influence action is 
zero.
In general, the influence functional
is a highly non--local object. Not only does it depend on the time 
history,
but --and this is the more important property-- it also
irreducibly mixes the two sets
of histories in the path integral of (3.5). 
In those cases where the initial decoupling
condition (3.4) is satisfied, the influence functional
depends only on the initial state of the environment. The influence 
functional
method can be extended to more general
conditions, such as thermal equilibrium between the system and the 
environment
\cite{HakAmb}, or correlated initial states \cite{FeyVer}.

The influence action for the open system (2.10), in a de Sitter 
phase, is derived in appendix A. It has the form
\begin{eqnarray}
S_{IF}[\phi,\phi']=&-&\int^t_{0}ds~\Delta(s)f(s) 
+\int_{0}^tds\int_{0}^sds'~
\Delta(s)\Sigma(s')\mu(s,s') 
\nonumber \\
&+&i\int_{0}^tds\int_{0}^s ds'~
\Delta(s)\Delta(s')\nu(s,s')
\end{eqnarray} 
where the sum and difference coordinates are
\begin{equation}
\Sigma(s)=\frac{1}{2}\Bigl(V''(\phi(s))+
V''(\phi'(s))\Bigr),\;\;\;
\Delta(s)=V''(\phi(s))-V''(\phi'(s)).
\end{equation}
The functions $\nu$ and $\mu$ are known respectively as the noise 
and dissipation kernels. This is because, as will be shown in 
section IV, 
they correspond to colored noise and non-local dissipation. The 
function $f(s)$ represents a renormalization of the potential.
These functions are derived exactly in appendix A. 

The approximate 
form of the noise kernel derived in (A44) is 
\begin{equation}
\nu(s,s')\simeq\frac{H^4 e^{6H\sigma}}{64\pi^6}
\cos\Bigl(4\epsilon\sinh(H\delta/2)\Bigr)+O(\epsilon^2),\;\;\;
\epsilon \ll 1\;{\rm and }\;\delta < \tau_c
\end{equation}
where
\begin{equation}
\delta=s-s',\;\;\;\sigma=(s+s')/2.
\end{equation}
As discussed in appendix A, $\tau_c$ is the correlation time of the noise
kernel which is 
defined as the time when the argument of the cosine 
equals $\pi/4$. The correlation timescale then becomes
\begin{equation}
\tau_c=2H^{-1}\ln(\pi/8\epsilon).
\end{equation}
The exact expression for the
dissipation kernel is derived in (A46). We do not quote
it here since it is a complicated expression that will play 
little further role. 
Inspection of (A41) and (A46), the exact noise and dissipation kernels, 
show that they quickly becomes
highly oscillatory when $\delta$ is greater than 
the correlation time $\tau_c$. This means the noise and dissipation 
kernels are effectively cutoff beyond the correlation time. 
This oscillatory behaviour is an artifact of the discrete low-frequency cutoff 
used for the bath.
From (A36) the potential renormalization term is
\begin{equation}
f(s)\simeq \frac{H^2}{8\pi^2}e^{3Hs}\Bigl(1/2+\ln(1/\epsilon)\Bigr),
\;\;\;\epsilon\ll 1.
\end{equation}
One of the most attractive features of this theory is that
the approximate noise and dissipation kernels (3.9) and (A48), 
along with (3.12) have no
algebraic dependence on $\epsilon$. This means that physical results 
will only very weakly depend on $\epsilon$.

\subsection{Local Approximation to the Influence Functional}
We can neglect the potential renormalization term in 
the influence action (3.7) when 
\begin{equation}
V(\phi)\gg \frac{H^2}{8\pi^2}V''(\phi)\ln(1/\epsilon).
\end{equation}
This holds when the potential satisfies the standard `slow roll' 
condition \cite{inflation}
\begin{equation}
|V''(\phi)|\ll 9H^2\simeq 24\pi V(\phi)/m_{pl}^2.
\end{equation}
It also holds more generally 
as long as $H\ll m_{pl}$. 

The slow roll conditions of inflation are necessary to ensure that 
an approximate de-Sitter inflation phase can proceed.
In the slow roll limit the dynamical timescale of $\phi$ is much 
longer than the Hubble timescale, so we expect to be able to make a 
local approximation
to the dynamics. We start by observing that
the integrals over $s'$ in the influence action (3.7) are strongly 
suppressed for $\delta > \tau_c$. Therefore 
the lower limit of the integral for $s'$ can be cutoff at $s-\tau_c$.
We can now invoke the slow roll condition. On the time scale $H^{-1}$
we expect $V''(\phi)$ to change little, so we can pull out 
$\Delta(s')$ and $\Sigma(s')$ from the integral over $s'$ in the 
influence action.
In this case the influence action becomes 
\begin{equation}
S_{IF}[\phi,\phi'] \simeq \int\limits_{0}^t ds~
\Delta(s)\Sigma(s)\int\limits_{s-\tau_c}^{s}ds'~\mu(s,s')+
i\int\limits_{0}^tds~\Delta^2(s)
 \int\limits_{s-\tau_c}^{s}ds'~\nu(s,s').       
\end{equation}

For the noise integral we now find
\begin{eqnarray}
\int\limits_{s-\tau_c}^{s}ds'\nu(s,s')&\simeq&
\frac{H^4e^{3Hs}}{64\pi^6}\int^s_{s-\tau_c}ds'~e^{3Hs'}
\nonumber \\
&\simeq& \frac{H^3e^{6Hs}}{192\pi^6}
\end{eqnarray}
where we used 
\begin{equation}
\cos x\simeq 1.
\end{equation}
For the dissipation kernel we use the result (A52) derived in appendix A 
which is
\begin{equation}
\int\limits_{s-\tau_c}^{s}ds'\mu(s,s')\sim 
\frac{e^{3Hs}}{4\pi^2}\ln(\pi/8\epsilon).
\end{equation}
The influence action (3.15) now becomes
\begin{equation}
S_{IF}[\phi,\phi'] \simeq 
\frac{\ln(\pi/8\epsilon)}{8 \pi^2}
\int\limits_{0}^tds~e^{3Hs}
\Bigl(V''^2(\phi(s))-V''^2(\phi'(s))\Bigr)+
\frac{iH^3}{192\pi^6}\int\limits_{0}^tds
~e^{6Hs}\Delta^2(s).      
\end{equation}

We see that the local approximation to the dissipation 
kernel has reduced to an effective potential term in 
the influence action. This term will be much smaller than the 
system potential when
\begin{equation}
V(\phi)\gg \ln(\pi/8\epsilon)\frac{V''^2(\phi)}{8\pi^2}.
\end{equation}
This will hold whenever the slow roll condition (3.14) is satisfied. 
As an example we can substitute $V=\lambda\phi^4$ into (3.20) which 
leads to the condition
\begin{equation}
\lambda\ll \frac{\pi^2}{18\ln(\pi/8\epsilon)}.
\end{equation}
This is clearly satisfied if $\lambda\ll 1$ which is the standard 
weak coupling condition.
Neglecting the potential term in (3.19), the  effective action 
becomes
\begin{equation}
S_{eff}[\phi,\phi']=S[\phi]-S[\phi']+
\frac{iH^3}{192\pi^6}\int^{t}_{0}
ds~e^{6Hs}\Delta^2(s)
\end{equation}
with $S[\phi]$ our scaled system action given by (A33) for the 
case of a de Sitter expansion phase. 

In general one must be careful in neglecting non-local dissipation 
in quantum open systems. Dissipation and noise, connected via a 
fluctuation-dissipation relation, are twin physical processes that 
reflect the relation between energy flowing into and out of the system.  
This relation is especially important when one is interested in possible
stationary states where a balance is eventually reached. In inflation 
we have a very flat potential well away from its minimum, and we are only 
interested in the dynamics over some relatively small finite time. 
As we showed above
it is these very special slow roll conditions of inflation that allowed
us to  neglect dissipation. This will certainly not be possible during 
the reheating phase but it should be a good first order approximation 
to the dynamics in the early slow roll phase.

\section{Stochastic Interpretation of the Influence Functional}
In this section we will show that a stochastic forcing term will 
generate the imaginary part of the influence action (3.7). Consider 
the action
\begin{equation}
S[\phi(s),\xi(s)]=\int_{t_i}^{t}ds~\Bigl(L(\phi,\dot{\phi},s)
+V''(\phi)\xi(s)\Bigr)
\end{equation}
where $\xi(s)$ is a zero-mean gaussian stochastic force. This system
generates the influence functional
\begin{equation}
{\cal F}[\phi,\phi']=\biggl\langle\exp\left[i\int_{t_i}^{t}
ds~\Delta(s)\xi(s)\right]\biggr\rangle,
\end{equation}
where the average is understood as a functional integral over 
$\xi(s)$, weighted
by a normalized gaussian probability density functional 
${\cal P}[\xi(s)]$.
The averaging can be performed to give \cite{HuMat1}
\begin{equation}
{\cal F}[\phi,\phi']=\exp\biggl\{
-\int\limits_{t_i}^tds\int\limits_{t_i}^{s}ds'
  ~\Delta(s)\Delta(s')\nu(s,s')\biggl\}
\end{equation}
where $\nu(s,s')$ is the second cumulant of the force $\xi$. Clearly 
then, before averaging over the noise, 
we can write our effective action derived from (3.7) as
\begin{equation}
S_{eff}[\phi,\phi',\xi]=S[\phi]-S[\phi']+\int_{t_i}^tds~
\Delta(s)\left[\int\limits_{t_i}^{s}ds'~
\Sigma(s')\mu(s,s')-f(s)+\xi(s)\right].
\end{equation}

This result can be used to show that  that the dynamics of the 
effective action (3.22), 
is equivalent to that generated by the 
quantum mechanical stochastic Hamiltonian 
\begin{equation}
H(t)=\frac{p^2}{2e^{3Ht}}+e^{3Ht}V(\phi)-\frac{H^{3/2}}{\sqrt{96}
\pi^3}e^{3Ht}~V''(\phi)F_w(t),
\end{equation}
where $p=e^{3Ht}\dot{\phi}$ is the canonical momentum and $p$ and 
$\phi$ obey the usual quantum mechanical commutator. 
$F_w(t)$ is a zero mean gaussian white noise with the correlation 
function
\begin{equation}
\langle F_w(t)F_w(t')\rangle=\delta(t-t').
\end{equation}
The white noise Hamiltonian (4.5) is an 
approximation to 
\begin{equation}
H(t)=\frac{p^2}{2e^{3Ht}}+e^{3Ht}V(\phi)-\frac{H^2}{8\pi^3}e^{3Ht}~
V''(\phi)F_c(t),
\end{equation}
where $F_c(t)$ is a zero mean gaussian colored noise of unit 
amplitude with the correlation function 
\begin{equation}
\langle F_c(t)F_c(t')\rangle=
\cos\Bigl(4\epsilon\sinh(H\delta/2)\Bigr).
\end{equation} 
This follows because the imaginary part of (3.22) is an 
approximation of the imaginary part of (3.7). 
Equations (4.5) and (4.7) are the main results of this paper. 
In the semiclassical limit the system timescale is clearly much greater 
than the Hubble time due to the slow rolling nature of inflation. In this 
case a white noise approximation should be reasonable. In the 
quantum regime it is not so clear that we can approximate (4.7) with 
(4.5). Numerical simulations will be required to answer this.

\subsection{Semiclassical Limit}
The semiclassical equation of motion for the general effective 
action (4.4) can be derived 
using
\begin{equation}
\frac{\delta \Bigl(S_{eff}[\phi,\phi',\xi]\Bigr)}
{\delta \Delta_{\phi}(t)}\bigg|_{\Delta_{\phi}
=0}=0,
\end{equation}
where $\Delta_{\phi}=\phi-\phi'$.
With the system action (A33) it can be shown that this 
equation becomes
\begin{equation}
\left(\ddot{\phi}+3\frac{\dot{a}}{a}\dot{\phi}+
V'(\phi)\right)a^3(t)+V'''(s)f(s)- 
V'''(\phi)\int_{t_i}^t \mu(t,s)V''(\phi(s))ds=
V'''(\phi)\xi(t),
\end{equation}
where $\langle \xi(t)\xi(t')\rangle=\nu(t,t')$.
Under special circumstances $\mu$ tends to the derivative of a delta 
function
which generates local dissipation, hence $\mu$ is referred to 
as the dissipation kernel. More generally
we see that in the semiclassical limit $\mu$ generates non-local 
dissipation. 
For the influence action (3.7), the functions $f$ and $\mu$ 
are unimportant, and  the 
semiclassical equation (4.10) becomes 
\begin{equation}
\ddot{\phi}+3H\dot{\phi}+V'(\phi)=\frac{H^2}{8\pi^3}V'''(\phi)F_c(t)
\end{equation}
where $F_c$ is the zero mean gaussian colored noise with correlator (4.8).
This equation is the semiclassical limit of the quantum open system whose 
dynamics is described by (4.7).
Invoking the slow roll approximation 
(neglecting the $\ddot{\phi}$) this equation becomes
\begin{equation}
\dot{\phi}+\frac{V'(\phi)}{3H}
=\frac{H}{24\pi^3}V'''(\phi)F_c(t).
\end{equation}

The neglect of the inertial term in (4.11) is standard practice in inflation 
when describing the dynamics of the mean field $\phi_m$. However it should also 
be 
a good approximation for describing the fluctuations $\delta\phi$ about the 
mean field ($\phi_m$ is the solution of (1.3)). 
Because these fluctuations are small we can linearize 
(4.11) about $\phi_m$ to obtain
\begin{equation}
\ddot{\delta\phi}+3H\dot{\delta\phi}+V''(\phi_m)\delta\phi
=\frac{H^2}{8\pi^3}V'''(\phi_m)F_c(t).
\end{equation}
The left hand side of this equation describes a damped simple harmonic 
oscillator. The standard over-damping condition is the same as the slow roll
condition (3.14). Therefore we can neglect the inertial term in 
(4.13) and conclude that (4.12) should also be a good description for the 
fluctuations as well as the mean field. 

The slow roll conditions allowed us to derive the local 
approximation (3.22) to the influence functional (3.7). 
In this approximation (4.12) becomes
\begin{equation}
\dot{\phi}+\frac{V'(\phi)}{3H}
=\frac{H^{1/2}}{\sqrt{864}\pi^3}V'''(\phi)F_w(t)
\end{equation}
where $F_w$ is a zero mean gaussian white noise with correlator (4.6).
This equation is the semiclassical limit of the result 
(4.5). It is a superior alternative to the 
Starobinsky equation (1.4) for reasons that are addressed 
in the discussion and conclusion section. 
Since this white noise equation is an 
approximation to the colored noise equation (4.12), it should be 
interpreted in the Stratonovich sense. 

We have motivated the 
relatively simple description (4.14) by invoking the slow roll 
conditions and assuming that a semiclassical description is valid. 
Ultimately the justification of equation (4.14) will require a detailed 
study of the quantum dynamics described by the Hamiltonian (4.7).
The complicated correlation function (4.8) would make this a  
difficult task. However it must be emphasized that the complex oscillatory
nature of this correlation function is purely an artifact of the discrete 
low frequency cutoff used for the bath. The only essential `invariant' 
information in (4.8) is the correlation time (3.11).
Therefore it would be reasonable to investigate the effects of colored 
noise by replacing the correlator (4.8) with 
an exponential correlation function with the correlation time 
(3.11). This correlator generates the simplest 
type of colored noise dynamics \cite{hugh} and would make the quantum 
dynamics described by (4.7) a tractable problem. 

\section{Amplitude of Density Fluctuations and the Fine Tuning 
Problem}
In this section we will use equation (4.14)
to estimate the amplitude of primordial density fluctuations 
generated by scalar field fluctuations in inflation. 
We will consider the potential  $V=\lambda\phi^4$, for which 
the solution to the slow roll equation (1.3) is
\begin{equation}
\phi(t)=\phi_0\exp\left(-\sqrt{\frac{2\lambda}{3\pi}}
m_{pl}~t\right).
\end{equation}
With this solution it can be shown that the two slow roll conditions 
of inflation \cite{inflation} break down 
for $\phi\sim m_{pl}$. We therefore require $\phi_0\gg m_{pl}$.
The number of e-folds in the inflationary era is given by 
\cite{inflation}
\begin{equation}
N(\phi)=\frac{8\pi}{m_{pl}^2}\int^{\phi_0}_{m_{pl}}d\phi~
\frac{V(\phi)}{V'(\phi)}\simeq \pi\frac{\phi_0^2}{m_{pl}^2}.
\end{equation}
A solution to the horizon problem requires $N\ge 60$. This means we 
require $\phi_0$ to be at least several times the Planck mass.

At this point it is useful to compare the dynamical timescale and 
the Hubble timescale. The dynamical timescale $\tau_d$ is found from 
the solution 
(5.1) to be
\begin{equation}
\tau_d=\sqrt{\frac{3\pi}{2\lambda}}m_{pl}^{-1}.
\end{equation}
Using 
\begin{equation}
H^2(\phi)=\frac{8\pi}{3m_{pl}^2}V(\phi)
\end{equation}
we find that 
\begin{equation}
\frac{\tau_d}{H^{-1}}=2\pi\phi^2/m_{pl}^2.
\end{equation}
When $\phi\gg m_{pl}$ the system timescale is much greater than the 
Hubble timescale. This provides a 
explicit example of the condition we assumed to make the local 
approximation to the influence functional.

So far we have only constrained $\phi_0$ 
to have a lower bound. In chaotic inflation we expect that initially
$V\sim m_{pl}^4$. This means that that we can only make 
$\phi_0\gg m_{pl}$ if we also correspondingly make $\lambda\ll 1$. 
In inflation it is the value of the density perturbations which 
serve to fix a value for $\lambda$, which by the potential 
constraint also fixes $\phi_0$.

The density contrast $\delta\rho/\rho$, that is generated by scalar 
field fluctuations $\delta\phi$, is approximately given by 
\cite{inflation,fluct} 
\begin{equation}
\frac{\delta\rho}{\rho}\simeq\frac{H}{\dot{\phi}}~\delta\phi.
\end{equation}
Assuming $\phi\gg\delta\phi$, we can linearize equation (4.14) to 
obtain
\begin{equation}
\dot{\delta\phi}+\frac{V''(\phi)}{3H}\delta\phi
=\frac{H^{1/2}}{\sqrt{864}\pi^3}V'''(\phi)F_w(t).
\end{equation}
We can further simplify this equation by putting $\phi\sim m_{pl}$ 
which 
corresponds to the end of inflation. From equation (5.4) the Hubble 
constant now becomes
\begin{equation}
H^2=\frac{8\pi}{3}\lambda m_{pl}^2.
\end{equation}
Equation (5.7) is now an Ornstein-Uhlenbeck process for which the 
stationary distribution gives
\begin{equation}
\langle\delta\phi^2\rangle=\frac{2\lambda^2m_{pl}^2}{9\pi^5}.
\end{equation}
In this case (5.6) becomes 
\begin{equation}
\frac{\delta\rho}{\rho}\sim \lambda.
\end{equation}
Observational constraints then require $\lambda\sim 10^{-6}$. 
Performing an exactly analogous calculation for the Starobinsky 
equation (1.4) gives
\begin{equation}
\frac{\delta\rho}{\rho} \sim \sqrt{\lambda}
\end{equation}
which leads to the conventional result $\lambda\sim 10^{-12}$. 

Clearly the theory developed here leads to a dramatic easing  
of the fine-tuning required in this model. 
The result (5.10) was obtained 
previously by Calzetta and Hu \cite{CH1} using an entirely 
different approach based on  
coarse-graining graviton degrees of freedom. They pointed out that 
this new value of $\lambda$ 
is consistent with the inflaton taking part in nonabelian gauge 
interactions with a coupling constant of $10^{-2}$.

\section{Discussion and Conclusion}
In this paper we have argued that the conventional theory of 
inflaton dynamics \cite{inflation}, along with Starobinsky's 
theory of stochastic inflation 
\cite{staro,stoinf}, were problematic because they could 
not explain the quantum-to-classical transition and they 
predicted an overproduction 
of primordial density fluctuations. It was concluded that a theory 
which could address these issues would involve a conceptual shift 
that comes with being based on 
the principles of non-equilibrium quantum statistical physics. 
There have been initial attempts to do this \cite{HuBelgium,sim}
using the system environment split originally proposed by 
Starobinsky \cite{staro}.
However, as pointed out in the introduction, these models are 
very complex and this has so far prevented 
them from being used in a detailed study of inflaton dynamics.

In this paper a new simplified theory of inflaton dynamics was 
developed. This approach is free of the problems associated 
with the conventional theory. 
It is based on the Hamiltonian (4.7)
which describes the quantum dynamics of 
the inflaton which has been coarse-grained over a constant proper 
volume of a de Sitter inflationary phase. The origin 
of the noise source is the backreaction of 
quantum fluctuations with wavelengths shorter than the 
coarse-graining scale. The noise is of a multiplicative nature 
because its origin is the mode-mode coupling 
induced by the self-interaction of the inflaton. For a free 
field this theory predicts no noise term as we should expect 
\cite{cgea,HuBelgium,morwas,CH1}. 
Apart from addressing the problems of the conventional theory, 
the other essential feature is the 
relative simplicity of 
the results (4.7) and (4.14). The major simplification was 
made by ignoring information about 
spatial correlations between the order parameters of different 
regions. This allows a description based on a single degree of 
freedom. Further significant simplification was obtained by 
invoking the standard slow roll assumptions. With this we were 
able to argue that the potential 
renormalization and non-local dissipation terms were negligible, 
and that the colored noise could be approximated by a white 
noise in the semiclassical limit. In this case 
the theory is independent of the arbitrary parameter $\epsilon$. 
The correlation time does depend on $\epsilon$ but only 
logarithmically. The 
robustness of this theory to the coarse-graining volume, which is 
parameterized by $\epsilon$, 
must be considered a major success. Another very important feature 
for applications to inflation is that the theory describes a 
minimally coupled scalar field with an arbitrary potential.

The slow roll semiclassical equation of motion (4.14), is a 
superior alternative to the Starobinsky equation for a number 
of reasons.
It is the approximate semiclassical limit of the quantum open 
system (4.7). 
This allows the validity of the semiclassical limit to be 
dynamically derived. 
The `classicality' of the Starobinsky equation is simply postulated 
and is not a natural outcome of any quantum open system. 
A quantum mechanical 
description, for which the Starobinsky equation is the 
semiclassical 
limit, has been attempted using methods developed some time ago to 
describe the quantum dynamics of classically dissipative systems 
\cite{graz}. 
However this amounts to simply `quantizing' an unjustified  
semiclassical equation of motion rather than 
deriving the quantum theory from fundamental physics.  
In any case, such a procedure is highly suspect because it
misconstrues the red-shift term $3H\dot{\phi}$ in the 
Starobinsky equation as damping. This repeats an old error in 
quantum mechanics of 
confusing a time dependent mass with a real dissipation 
\cite{error}. 
The other major advantage is that equation (4.14) leads to a 
dramatic easing of the 
fine tuning required in the theory.
We showed this for the $\lambda\phi^4$ potential. 
The fine tuning problem is systematic to all potentials 
of interest in inflation. Similarly, we should expect that 
equation (4.14) 
will lead to a large easing of the fine tuning required for 
these potentials. 
Finally, the dramatic easing of the fine tuning constraints will 
make equation (4.14) much simpler to numerically simulate 
than the Starobinsky equation. Yi and Vishniac (1993) 
\cite{sifluct} 
numerically simulated the Starobinsky equation for the 
$\lambda\phi^4$ potential. They used $\lambda=10^{-6}$ in their 
simulations because the supposedly correct value of $10^{-12}$
was numerically intractable. The former value of $\lambda$ 
is a natural outcome of the theory presented here. 

As for future work, the quantum decoherence generated by the noise 
term in (4.7) will allow a dynamical investigation of entropy generation 
and the 
quantum-to-classical transition. Decoherence studies have 
been previously attempted \cite{HuBelgium,sim}. However, 
because these models were very complicated the authors were 
restricted to numerically calculating the diffusion coefficient 
in the quantum master equation. Unfortunately this tells us 
nothing about whether or not enough decoherence will occur in 
the model to justify a semiclassical description. The critical 
dynamical aspect to this problem is the competition between the 
coherence generating non-linear potential and inflation expansion 
effects, and the decohering stochastic term. 
This can only be investigated with a full solution of the 
non-unitary quantum dynamics.
If the noise term is strong enough we would hope that 
decoherence will justify a semiclassical description. 
However, we must also remember that the noise can not be so 
large as to destroy the essentially deterministic 
slow roll dynamics of the local order parameter $\phi$. The 
inflaton dynamics described by equation (4.7) is the first 
viable theory to permit a detailed study of the 
quantum-to-classical of the local order parameter $\phi$. 
This is the critical process that leads to the generation 
of classical density fluctuations.

Assuming that the semiclassical description (4.14) is valid, we 
can use this equation to 
deduce the amplitude and spectrum of the generated primordial 
density fluctuations, along with the implications this has for 
coupling constants of various inflaton potentials. 
Also of great interest is the generation of any relevant 
non-gaussian features in the resulting probability distribution 
of $\phi$ \cite{sifluct}, and the implications of equation (4.14) 
for the very large scale structure of the 
universe \cite{vlss}. Till
now, these issues have been addressed with the Starobinsky 
equation as their theoretical foundation. 
An interesting technical issue is what effect relaxing the white 
noise approximation would have on these results. 
Investigations into the implications of the theory presented here 
are in progress. 

\acknowledgments
I would like to thank the Australia Research Council for their 
generous support of this research 
through an Australian Postdoctoral Research Fellowship and an ARC 
small grant.
 
\appendix
\section{Calculating the Influence Functional}
The exact influence functional for the action of the type 
(2.10) has been previously found to be \cite{HuMat1}
\begin{equation}
{\cal F}[\phi,\phi']=\prod_k\Bigl\{\alpha_k
[\phi']\alpha_k^*[\phi]-\beta_k[\phi']\beta_k^*[\phi]\Bigr\}^{-1},
\end{equation}
where $\alpha,\beta$ are Bogolubov coefficients which 
must satisfy the constraint 
\begin{equation}
|\alpha|^2-|\beta|^2=1.
\end{equation}
The influence functional (A1) differs from that in \cite{HuMat1} 
because we have here 
taken into account the double set of modes in the action (2.10).
The Bogolubov coefficients are solutions of the 
equations
\begin{eqnarray}
\dot{\alpha} & = & -ig\beta-ih\alpha \\
\dot{\beta} & = & ih\beta+ig\alpha
\end{eqnarray}
where
\begin{equation}
g(s)=\frac{1}{2}\left(\frac{a^3(s)\omega^2(s)}{k}
-\frac{k}{a^3(s)}\right)
\end{equation}
\begin{equation}
h(s)=\frac{1}{2}\left(\frac{k}{a^3(s)}
+\frac{a^3(s)\omega^2(s)}{k}\right),
\end{equation}
and
\begin{equation}
\omega^2(s)=k^2/a^2+V''(\phi(s)).
\end{equation}
We must have $\alpha=1$ and $\beta=0$ at $t_i$ as
our initial conditions.

Defining
\begin{equation}
{\bf U}(t,t_i)= \left(\begin{array}{cc}
\alpha [\phi_{t,t_i}] & \beta^* [\phi_{t,t_i}] \\
\beta [\phi_{t,t_i}] & \alpha^* [\phi_{t,t_i}] \end{array} \right),
\end{equation}
the solution of (A3-4) can be written as
\begin{equation}
{\bf U}(t,t_i)={\cal T}\exp\left(-i\int_{t_i}^tds~{\bf u}(s)\right)
\end{equation}
where
\begin{equation}
{\bf u}(s)=
\left(\begin{array}{cc} 
h(s) & g(s) \\ -g(s)&-h(s) 
\end{array}\right). 
\end{equation}
We will write
\begin{equation}
{\bf u(s)}={\bf u_0(s)}+{\bf u_1(s)}
\end{equation}
where
\begin{equation}
{\bf u_1}(s)=\frac{a^3(s)V''(\phi(s))}{2k}\left(
\begin{array}{cc}
1 & 1 \\
-1 & -1
\end{array} \right).
\end{equation}
In this case we have 
\begin{equation}
{\bf U}(t,t_i)={\bf U}_0(t,t_i){\bf U}_1(t,t_i)
\end{equation} 
where ${\bf U}_0(t,t_i)$ is ${\bf U}(t,t_i)$ evaluated at 
$V''(\phi)=0$.
The equation of motion for ${\bf U}_1$  is
\begin{equation}
\dot{{\bf U}}_1(s,t_i)=-i\Bigl({\bf U}_0^{-1}(s,t_i){\bf u}_1(s)
{\bf U}_0(s,t_i)\Bigr){\bf U}_1(s,t_i)
\end{equation}
where ${\bf U}^{-1}$ denotes the inverse matrix. This 
inverse is
\begin{equation}
{\bf U}^{-1}(t,t_i)= \left(\begin{array}{cc}
\alpha^* [\phi_{t,t_i}] & -\beta^* [\phi_{t,t_i}] \\
-\beta [\phi_{t,t_i}] & \alpha [\phi_{t,t_i}] \end{array} \right)
\end{equation}
because of the constraint (A2). From (A13) and (A8) we have 
\begin{equation}
\alpha [\phi_{t,t_i}]=\alpha_0(t,t_i)\alpha_1 [\phi_{t,t_i}]+
\beta_0^*(t,t_i)\beta_1 [\phi_{t,t_i}]
\end{equation}
\begin{equation}
\beta [\phi_{t,t_i}]=\beta_0(t,t_i)\alpha_1 [\phi_{t,t_i}]+
\alpha_0^*(t,t_i)\beta_1 [\phi_{t,t_i}].
\end{equation}
Substituting this into the influence functional 
(A1), and using the constraint (A2),
we find 
\begin{equation}
{\cal F}[\phi,\phi']=\prod_k\Bigl\{\alpha_1
[\phi']\alpha_1^*[\phi]-\beta_1[\phi']\beta_1^*[\phi]\Bigr\}^{-1}
\end{equation}
where we have dropped the $k$ subscript on the 
Bogolubov coefficients.

To second order in $V''$ the solution to (A14) is 
\begin{eqnarray}
{\bf U}_1(t,t_i)&=&
1-i\int_{t_i}^t ds~{\bf U}_0^{-1}(s,t_i){\bf u}_1(s)
{\bf U}_0(s,t_i) \nonumber \\
&-& \int_{t_i}^tds\int_{t_i}^s ds'~{\bf U}_0^{-1}(s,t_i){\bf u}_1(s)
{\bf U}_0(s,t_i){\bf U}_0^{-1}(s',t_i){\bf u}_1(s')
{\bf U}_0(s',t_i).
\end{eqnarray}
Using this solution and (A18), the influence action becomes
\begin{eqnarray}
S_{IF}[\phi,\phi']&=&
\sum_k\biggl\{-\frac{1}{2k}\int^t_{t_i}ds~a^3(s)\Delta(s)
X_k(s)X_k^*(s) \nonumber \\
&+&\frac{i}{8k^2}\int_{t_i}^tds\int_{t_i}^s ds'~a^3(s)a^3(s')
\Delta(s)\Delta(s')
\Bigl[X^{*2}_k(s)X^{2}_k(s')+X^{2}_k(s)X^{*2}_k(s')\Bigr] 
\nonumber \\
&-&\frac{i}{4k^2}\int_{t_i}^tds\int_{t_i}^sds'~a^3(s)a^3(s')
\Delta(s)\Sigma(s')
\Bigl[X^{*2}_k(s)X^{2}_k(s')-X^{2}_k(s)X^{*2}_k(s')\Bigr] \biggr\}
\end{eqnarray}
where 
\begin{equation}
\Delta(s)=V''(\phi(s))-V''(\phi'(s)),\;\;\;
\Sigma(s)=\frac{1}{2}\Bigl[V''(\phi(s))+V''(\phi'(s))\Bigr]
\end{equation}
and 
\begin{equation}
X_k(s)=\alpha_{0}(s,t_i)+\beta_{0}(s,t_i).
\end{equation}
Using (A3-4) we can show that (A22) obeys the classical equation of 
motion 
\begin{equation}
\ddot{X}+3\frac{\dot{a}}{a}\dot{X}+\frac{k^2}{a^2}X=0
\end{equation}
subject to the initial conditions
\begin{equation}
X(t_i)=1,\;\;\; \dot{X}(t_i)=-ik.
\end{equation}
These initial conditions ensure $\alpha(t_i)=1$ and $\beta(t_i)=0$.

Taking the continuum limit  we have 
\begin{equation}
\sum_k\rightarrow\frac{\Omega}{4\pi^2}\int^{\infty}_{k_{min}}
k^2dk.
\end{equation}
where we adopt the notation that $\Omega$ is a symmetric function of 
$s$ and $s'$, 
i.e. $\Omega^2\equiv\Omega(s)\Omega(s')$.
In this case the influence action (A20) becomes 
\begin{eqnarray}
S_{IF}[\phi,\phi']=&-&\int^t_{t_i}ds~\Delta(s)f(s) 
+\int_{t_i}^tds\int_{t_i}^sds'~
\Delta(s)\Sigma(s')\mu(s,s') 
\nonumber \\
&+&i\int_{t_i}^tds\int_{t_i}^s ds'~
\Delta(s)\Delta(s')\nu(s,s')
\end{eqnarray}
where
\begin{equation}
f(s)=a^3(s)\frac{\Omega}{8\pi^2}\int^{\infty}_{k_{min}}
dk~kX_k(s)X_k^*(s),
\end{equation}
\begin{equation}
\nu(s,s')=a^3(s)a^3(s')\frac{\Omega}{32\pi^2}\int^{\infty}_{k_{min}}
dk~\Bigl[X^{*2}_k(s)X^{2}_k(s')+X^{2}_k(s)X^{*2}_k(s')\Bigr]
\end{equation}
and
\begin{equation}
\mu(s,s')=-ia^3(s)a^3(s')\frac{\Omega}{16\pi^2}
\int^{\infty}_{k_{min}}dk~
\Bigl[X^{*2}_k(s)X^{2}_k(s')-X^{2}_k(s)X^{*2}_k(s')\Bigr].
\end{equation}

From the action (2.8) we see that it is $\psi^2$ that couples 
to the system.
The truncation of the solution (A19) amounts to assuming that 
$\psi^2$ generates a gaussian noise source on the system. This 
is proved in section 4. The gaussian property ensures that only 
the first two cumulants of the stochastic process generated by 
$\psi^2$ appears in the influence functional. 
The influence functional can be expressed in the alternative form 
\cite{HuMat1} 
\begin{eqnarray}
f(s)&\sim&\langle\psi^2(s)\rangle \nonumber \\ 
\mu(s,s')&\sim&\langle\psi^2(s)\psi^2(s')\rangle-\langle\psi^2(s')
\psi^2(s)\rangle \nonumber  \\
\nu(s,s')&\sim& \langle\psi^2(s)\psi^2(s')\rangle+\langle\psi^2(s')
\psi^2(s)\rangle-2\langle\psi^2(s)\rangle\langle\psi^2(s')\rangle
\end{eqnarray}
which makes this clearer. For a gaussian process all higher 
order cumulants vanish because all higher order 
moments can be expressed in terms of first and second moments. 
Clearly then, the truncation of the solution (A19) is an
equivalent approximation to writing all higher order moments of 
$\psi^2$ in terms of first and second moments.

Our quantum field theory has the commutator
\begin{equation}
\Bigl[\Phi({\bf x}),P_{\Phi}({\bf y})\Bigr]=
i\hbar\delta({\bf x}-{\bf y})
\end{equation}
where $P_{\Phi}({\bf y})$ is the canonical field momentum 
derived from the action (2.3). Using this commutator 
we can show that our coarse-grained 
field (2.5), and a similarly defined  coarse-grained canonical 
momentum, 
will obey the usual single particle quantum 
mechanical commutator only after we perform the scaling
\begin{equation}
\hbar\rightarrow \Omega(s)\hbar
\end{equation}
on the Planck constant. This is a natural consequence 
of reducing a quantum field theoretic problem to a 
quantum mechanical problem and has been discussed 
previously \cite{graz}. We have set $\hbar=1$ in this paper, but the 
scaling (A32) is equivalent to 
scaling the effective action (4.4) by $\Omega(s)$. This gives the 
new system action 
\begin{equation}
S[\phi]=\int^t_{t_i}ds~a^3(s)
\left[\frac{1}{2}\dot{\phi}^2(s)-V(\phi)\right]
\end{equation}
and it requires that we must scale the functions (A27-A29) as 
\begin{equation}
f(s)\rightarrow\Omega^{-1}f(s),\;\;\;
\mu(s,s')\rightarrow\Omega^{-1}\mu(s,s'),\;\;\;
\nu(s,s')\rightarrow\Omega^{-2}\nu(s,s').
\end{equation}

Equation (A23) is identical to (2.13) when we scale $X$ by the scale 
factor and transform to conformal time. 
For a de Sitter phase, the solution of (2.13) leads to  
\begin{equation}
X_k(s)= \frac{e^{-ik\eta}}{a(s)}\left(1-\frac{i}{k\eta}\right).
\end{equation}
Strictly speaking an appropriate linear combination 
of these complex mode functions are required to satisfy the initial 
conditions (A24). However the time dependent 
low frequency cutoff ensures that any dependence on 
the initial condition is a transient effect. 
So the description based on the mode function (A35) is accurate on 
the Hubble timescale.
With (A35) we are now in a position to calculate (A27-A29) using (2.15) 
as the low frequency cutoff. 

We will first consider the function $f(s)$. 
This function is ultraviolet divergent which
is not surprising since $f\sim\langle\psi^2\rangle$.
We will adopt the usual non-rigorous 
renormalization procedure here simply by imposing the  
inflationary ultraviolet cutoff $k=Ha$ (see Brandenberger (1984) 
\cite{fluct} and Habib \cite{Hab}). This 
cutoff is determined by demanding that the energy density of 
inflaton fluctuations be less than the potential (which is supposed to 
dominate), and that the inflaton fluctuations can be taken to be in 
the vacuum state. This is discussed by Liddle and Lyth 
\cite{inflation}. 
We must also perform the scaling (A34) 
on (A27) which eliminates the volume factor $\Omega$ from (A27). We 
therefore find
\begin{eqnarray}
f(s)&=&\frac{a(s)}{8\pi^2}\int^{Ha}_{\epsilon Ha}
~dk\left(k+\frac{1}{k\eta^2}\right) \nonumber \\ 
&=& \frac{H^2}{8\pi^2}e^{3Hs}\Bigl(1/2+\ln(1/\epsilon)\Bigr),
\;\;\;\epsilon\ll 1.
\end{eqnarray}
The small $\epsilon$ limit is necessary in order to 
ignore the spatial gradient term in the system sector. 
This was shown in section 2 where it was also discussed that we require 
$\epsilon$ to be small but {\it finite}. We are therefore never faced 
with any problems associated with infrared divergences.

After performing the scaling 
(A34) on (A28) and substituting (A35) into (A28) we find that the 
noise kernel becomes
\begin{eqnarray}
\nu(s,s')&=&\frac{a(s)a(s')}{16\Omega\pi^2}\int_{\epsilon Ha}^{\infty}~dk
\left[\cos 2k\delta_{\eta}\left(1+\frac{2}{k^2\eta\eta'}-\frac{\delta_{\eta}^2}
{k^2\eta^2\eta'^2}+\frac{1}{k^4\eta^2\eta'^2}\right)\right. \nonumber \\
&+& \left. 2\delta_{\eta}\sin 2k\delta_{\eta}\left(\frac{1}{k\eta\eta'}+
\frac{1}{k^3\eta^2\eta'^2}\right)\right]
\end{eqnarray}
where 
\begin{equation}
\delta_{\eta}=\eta-\eta'.
\end{equation}
The noise kernel 
is clearly ultraviolet {\it finite} and therefore requires no 
renormalization.  
This should not be surprising. From (A30) 
we know that the kernels are built from the quadratic field 
operators $\psi^2$ rather than the linear operators $\psi$
of conventional two-point functions (which are 
well known to be ultraviolet divergent).
The integrals in (A37) can be done by using the integral identities
\begin{equation}
\int\frac{e^{ikx}}{k^m}~dk=\frac{1}{m-1}\left[-\frac{e^{ikx}}{k^{m-1}}
+ix\int\frac{e^{ikx}}{k^{m-1}}~dk\right]
\end{equation}
\begin{equation}
\int\frac{e^{ikx}}{k}~dk={\rm Ci}(kx)+i{\rm Si}(kx)
\end{equation}
where ${\rm Si}$ and ${\rm Ci}$ are the sine and cosine integral 
functions .
We then find that (A37) becomes
\begin{eqnarray}
\nu(s,s')&=&\frac{H^4}{128\pi^6}e^{6H\sigma}
\left[3\pi\epsilon^3 H^{-1}\delta_d(\delta) 
+\cos\Bigl(4\epsilon\sinh(H\delta/2)\Bigr)\Bigl(2+12\epsilon^2+
8\epsilon^2\sinh^2(H\delta/2)\Bigr) \right. \nonumber \\
&+&
\sin\Bigl(4\epsilon\sinh(H\delta/2)\Bigr)
\left(8\epsilon\sinh(H\delta/2)-\frac{3\epsilon^3}{2}\sinh^{-1}(H\delta/2)
\right) \nonumber \\
&+& \left. 16\epsilon^3 {\rm Si}\Bigl(4\epsilon\sinh(H\delta/2)\Bigr)
\Bigl(3\sinh(H\delta/2)+2\sinh^3(H\delta/2)\Bigr)\right],
\end{eqnarray}
where $\delta_d$ denotes the Dirac delta function which is not to be confused 
with $\delta$ defined in (3.10).
This noise kernel does not diverge for $\epsilon\ll 1$
because the scaling (A34) makes the kernel proportional to $\Omega^{-1}$.
Using (2.11) with $\Lambda=Ha$ we find that 
\begin{equation}
\Omega^{-1}=\frac{3\epsilon^3H^3e^{3Hs}}{4\pi^4}
\end{equation}
which shows that $\Omega^{-1}$ scales the kernel by $\epsilon^3$.
The noise kernel (A41) has a correlation time which we 
will define as the time when the argument of the cosine in (A41)
equals $\pi/4$. We then find that the correlation timescale 
$\tau_c$ is
\begin{equation}
\tau_c=2H^{-1}\ln(\pi/8\epsilon).
\end{equation}
The noise kernel is highly oscillatory for 
$\delta > \tau_c$. It is therefore effectively cutoff beyond 
the correlation time. We can therefore consider the approximate 
noise kernel
\begin{equation}
\nu(s,s')\simeq \frac{H^4}{64\pi^6}e^{6H\sigma}
\cos\Bigl(4\epsilon\sinh(H\delta/2)\Bigr)+O(\epsilon^2),\;\;\;
\epsilon\ll 1 \;{\rm and}\;\delta<\tau_c.
\end{equation}
Clearly the noise kernel has simplified greatly. However what is most 
pleasing is that the noise kernel has lost its algebraic 
dependence on $\epsilon$. Because the noise kernel is ultraviolet 
finite we didn't impose the inflationary cutoff. Had we 
done so the cutoff would have contributed terms of order $\epsilon^3$ 
into the noise kernel. Thus in the small $\epsilon$ limit the noise kernel 
is also independent of any ultraviolet cutoff.

We will now consider the dissipation kernel (A29). The scaling (A34)
eliminates the volume factor $\Omega$ from the kernel in (A29) which upon 
substituting (A35) becomes
\begin{eqnarray}
\mu(s,s')&=&\frac{a(s)a(s')}{8\pi^2}\int_{\epsilon Ha}^{Ha}~dk
\left[\sin 2k\delta_{\eta}\left(1+\frac{2}{k^2\eta\eta'}-\frac{\delta_{\eta}^2}
{k^2\eta^2\eta'^2}+\frac{1}{k^4\eta^2\eta'^2}\right)\right. \nonumber \\
&-& \left. 2\delta_{\eta}\cos 2k\delta_{\eta}\left(\frac{1}{k\eta\eta'}+
\frac{1}{k^3\eta^2\eta'^2}\right)\right].
\end{eqnarray}
All of the terms in the above integral are ultraviolet finite except for the 
first flat space contribution which leads to a logarithmic divergence. 
For this reason we have used the inflationary ultraviolet cutoff.
Using the integrals (A39-40) we find that (A45) becomes
\begin{eqnarray}
\mu(s,s')&=&\frac{H}{48\pi^2}e^{3H\sigma}
\left[\sin\Bigl(4\epsilon\sinh(H\delta/2)\Bigr)\Bigl(2\epsilon^{-3}
+12\epsilon^{-1}+8\epsilon^{-1}\sinh^2(H\delta/2)\Bigr) \right. \nonumber \\
&+&
\cos\Bigl(4\epsilon\sinh(H\delta/2)\Bigr)
\left(\frac{3}{2}\sinh^{-1}(H\delta/2)-8\epsilon^{-2}\sinh(H\delta/2)
\right) \nonumber \\
&-&  16 {\rm Ci}\Bigl(4\epsilon\sinh(H\delta/2)\Bigr)
\Bigl(3\sinh(H\delta/2)+2\sinh^3(H\delta/2)\Bigr) \nonumber \\
&-&\sin\Bigl(4\sinh(H\delta/2)\Bigr)\Bigl(14+
8\sinh^2(H\delta/2)\Bigr)  \nonumber \\
&-&\cos\Bigl(4\sinh(H\delta/2)\Bigr)
\left(\frac{3}{2}\sinh^{-1}(H\delta/2)-8\sinh(H\delta/2)
\right) \nonumber \\
&+& \left. 16 {\rm Ci}\Bigl(4\sinh(H\delta/2)\Bigr)
\Bigl(3\sinh(H\delta/2)+2\sinh^3(H\delta/2)\Bigr)\right].
\end{eqnarray}
We are only interested in the dissipation kernel for $\delta <\tau_c$. 
For times greater than this the dissipation kernel is highly oscillatory 
and therefore effectively cutoff. In this regime we can use the 
approximations 
\begin{equation}
\cos x\simeq 1-x^2/2,\;\;\;\sin x\simeq x-x^3/3
\end{equation}
to get the leading order approximation to the dissipation kernel (A46) 
which is
\begin{eqnarray}
\mu(s,s')&\simeq&\frac{H}{\pi^2}e^{3H\sigma}
\left[\frac{1}{32}\sinh^{-1}(H\delta/2)
+\sinh(H\delta/2)+\frac{10}{9}\sinh^3(H\delta/2)
\right. \nonumber \\
&-&  {\rm Ci}(4\epsilon\sinh(H\delta/2))
\Bigl(\sinh(H\delta/2)+\frac{2}{3}\sinh^3(H\delta/2)\Bigr) \nonumber \\
&-&\sin\Bigl(4\sinh(H\delta/2)\Bigr)\Bigl(\frac{7}{24}+
\frac{1}{6}\sinh^2(H\delta/2)\Bigr)  \nonumber \\
&-&\cos\Bigl(4\sinh(H\delta/2)\Bigr)
\left(\frac{1}{32}\sinh^{-1}(H\delta/2)-\frac{1}{6}\sinh(H\delta/2)
\right)  \\
&+& \left. {\rm Ci}\Bigl(4\sinh(H\delta/2)\Bigr)
\Bigl(\sinh(H\delta/2)+\frac{2}{3}\sinh^3(H\delta/2)\Bigr)
+O(\epsilon^2)\right],\;\;\;\epsilon\ll 1\;{\rm and}\;\delta <\tau_c. 
\nonumber
\end{eqnarray}
Remarkably we find that in this approximation the dissipation kernel has lost 
its algebraic dependence on $\epsilon$. 

Unfortunately the dissipation kernel can not be approximated to a simple 
expression like the noise kernel. However the slow roll conditions of 
inflation ensure that we need only consider an order of magnitude estimate of 
the integral 
\begin{equation}
\int^s_{s-\tau_c}~ds'\mu(s,s')\sim \tau_c~\mu(s,s-\tau_c)
\end{equation}
which we obtain by multiplying the correlation time $\tau_c$ defined in 
(A43) by the 
dissipation kernel evaluated at $\delta=\tau_c$.
From (A48) we can write
\begin{equation}
\mu(s,s-\tau_c)\simeq\frac{10H}{9\pi^2}e^{3Hs}e^{-3H\tau_c/2}
\sinh^3(H\tau_c/2)\simeq\frac{H}{8\pi^2}e^{3Hs}
\end{equation}
where we used 
\begin{equation}
\sinh (H\tau_c/2)\simeq e^{H\tau_c/2}/2.
\end{equation}
Using (A43) and (A50) we find that (A49) becomes
\begin{equation}
\int^s_{s-\tau_c}~ds'\mu(s,s')\sim\frac{e^{3Hs}}{4\pi^2}\ln(\pi/8\epsilon).
\end{equation}

\end{document}